\documentclass[twocolumn,footinbib,showpacs,preprintnumbers,amsmath,amssymb,superscriptaddress]{revtex4-1}

\usepackage{graphicx}
\usepackage{dcolumn}
\usepackage{bm}

\begin{document}

\title{Pauli paramagnetism of an ideal Fermi gas}

\author{Ye-Ryoung Lee}
\affiliation{MIT-Harvard Center for Ultracold Atoms, Research Laboratory of Electronics, Department of Physics, Massachusetts Institute of Technology, Cambridge, MA 02139, USA}
\author{Tout T. Wang}
\affiliation{MIT-Harvard Center for Ultracold Atoms, Research Laboratory of Electronics, Department of Physics, Massachusetts Institute of Technology, Cambridge, MA 02139, USA}
\affiliation{Department of Physics, Harvard University, Cambridge, MA 02138, USA}
\author{Timur M. Rvachov}
\affiliation{MIT-Harvard Center for Ultracold Atoms, Research Laboratory of Electronics, Department of Physics, Massachusetts Institute of Technology, Cambridge, MA 02139, USA}
\author{Jae-Hoon Choi}
\affiliation{MIT-Harvard Center for Ultracold Atoms, Research Laboratory of Electronics, Department of Physics, Massachusetts Institute of Technology, Cambridge, MA 02139, USA}
\author{Wolfgang Ketterle}
\affiliation{MIT-Harvard Center for Ultracold Atoms, Research Laboratory of Electronics, Department of Physics, Massachusetts Institute of Technology, Cambridge, MA 02139, USA}
\author{Myoung-Sun Heo}
\altaffiliation[Permanent address: ]{Korea Research Institute of Standards and Science, Daejeon 305-340, Korea}
\affiliation{MIT-Harvard Center for Ultracold Atoms, Research Laboratory of Electronics, Department of Physics, Massachusetts Institute of Technology, Cambridge, MA 02139, USA}


\begin{abstract}
We show how to use trapped ultracold atoms to measure the magnetic susceptibility of a two-component Fermi gas.  The method is illustrated for a non-interacting gas of $^6$Li, using the tunability of interactions around a wide Feshbach resonances. The susceptibility versus effective magnetic field is directly obtained from the inhomogeneous density profile of the trapped atomic cloud. The wings of the cloud realize the high field limit where the polarization approaches 100\%, which is not accessible for an electron gas.
\end{abstract}

\pacs{67.85.Lm, 67.10.Db}
\maketitle

\section{Introduction}
Ultracold atoms can be prepared with almost complete control over their density, temperature and interactions. They serve as model systems for exploring unsolved problems in many-body physics as well as for demonstrating well-known textbook physics, such as of ideal, non-interacting gases that do not exist in nature. For example several phenomena related to Pauli blocking of fermions were clearly demonstrated only after the advent of ultracold Fermi gases \cite{ketterleVarennaFermi}. Here we use the tunability of atomic interactions near Feshbach resonances to create a non-interacting Fermi gas with two components. This is
a realization of an ideal, non-interacting free electron gas (FEG) with spin up and down components, as assumed in the simple theory of metals. We demonstrate how such an ideal Fermi gas will respond to effective magnetic fields, which is described by Pauli paramagnetism.

The paper is mainly pedagogical. It explains how paramagnetism is observed in trapped atomic samples which have an inhomogeneous density due to the harmonic confinement potential. Furthermore, these atomic samples realize a canonical ensemble in the fixed atom numbers $N_\uparrow$ and $N_\downarrow$ of the two components, whereas in metals and solid-state physics, $N_\uparrow - N_\downarrow$ is the magnetization determined by the applied external magnetic field.  In all previous studies of paramagnetism, the magnetization was weak since the applied magnetic field times the magnetic moment was much smaller than the Fermi energy \cite{SchumacherLi,SchumacherNa,Schultz,Kaeck}.  With ultracold atoms, we can easily realize the strong field case where the chemical potential difference is larger than the Fermi energy and therefore fully polarizes the gas. Besides its pedagogical purpose, this paper experimentally demonstrates Pauli paramagnetism in a truly non-interacting system exactly described by basic theory, whereas measurements even in simple metals revealed major discrepancies due to interaction effects \cite{SchumacherLi,SchumacherNa,Schultz,Kaeck}.

\section{Concept}\label{sec:concept}
\begin{figure}
  \includegraphics[scale=0.6]{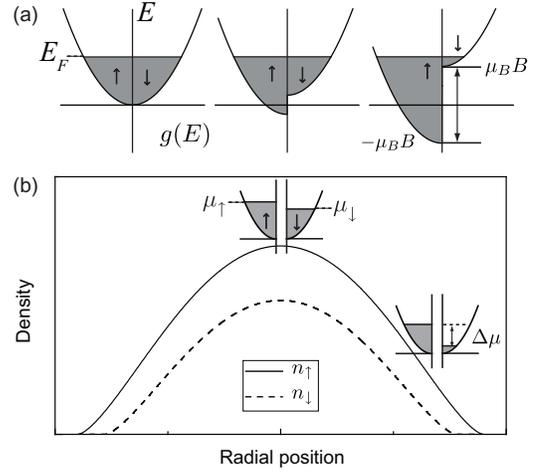}\\
  \caption{Comparison between a FEG and a trapped two-component Fermi gas at temperature $T=0$. (a) Response of the FEG to an external magnetic field $B$. Electrons in the two spin states are filled up to the Fermi energy $E_F$. An external magnetic field $B$ shifts the energy of the two states into opposite directions and produces the density difference, or magnetization. (b) Density distributions $n_\uparrow$ and $n_\downarrow$ of a two-component trapped Fermi gas at $T=0$. As discussed in the text, the local density polarization and corresponding normalized effective magnetic field field varies across the atomic cloud. This is illustrated in the diagrams next to the curve for   $n_\uparrow$, where the gaps between the two spin states represent the absence of spin-flip processes in the system mentioned in the text.}\label{fig:paramagnetism}
\end{figure}

\begin{figure}[h]
  \includegraphics[scale=0.8]{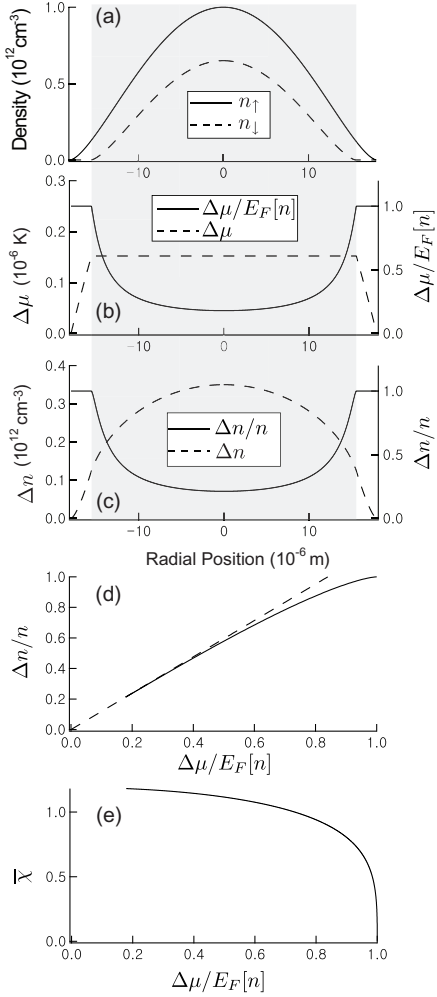}\\
  \caption{Determination of the susceptibility $\overline{\chi}$ from atomic density profiles. (a) Simulated density profiles of a two-component Fermi gas in a spherically symmetric harmonic trap at zero temperature. The total population imbalance $\delta\equiv(N_\uparrow-N_\downarrow)/(N_\uparrow+N_\downarrow)=21\%$. The shaded area corresponds to the partially polarized region where $\overline{\chi}$ is well-defined. (b) The local chemical potential difference $\Delta\mu$ with (solid) and without (dashed) normalization by the Fermi energy $E_F[n]$. For the fully polarized gas $\overline{\Delta \mu}=\Delta\mu/E_F[n]=1$ according to the definition in the text. (c) The density difference $\Delta n$ with (solid) and without (dashed) normalization by the total density $n$. (d) Normalized density difference $\overline{\Delta n}$ vs. normalized chemical potential difference $\overline{\Delta \mu}$. For the fully polarized gas, $\overline{\Delta n}=\overline{\Delta \mu}=1$. A linear fit at low $\overline{\Delta \mu}$ (dashed line) gives $\overline{\chi}=3/2^{4/3}\approx1.19$. (e) Normalized magnetic susceptibility $\overline{\chi}$ versus the normalized chemical potential difference $\overline{\Delta \mu}$.}\label{fig:simulation}
\end{figure}

Pauli paramagnetism explains the magnetization of a free electron gas with two spin states, neglecting the contribution from orbital motion. In a FEG, the applied field shifts the energies of the two spin states in opposite directions as in Fig.~\ref{fig:paramagnetism}(a). For a uniform FEG at temperature $T$=0, the total number $N_{\uparrow(\downarrow)}$ of electrons in each spin state inside a volume $V$ is given by
\begin{equation}\label{eq:solid_state}
N_{\uparrow(\downarrow)}=\int_{\mp\mu_B B}^{E_F}g(E\pm\mu_B B)dE,
\end{equation}
where $g(E)=\frac{V}{4\pi^2}\left(\frac{2m}{\hbar^2}\right)^{3/2}E^{1/2}$ is the density of states,  $E_F$ the Fermi energy, $m$ the mass of an electron, $\mu_B$ the Bohr magneton and $\hbar$ Planck's constant. The magnetic susceptibility $\chi$ is defined as $\mu_B\partial\Delta n/\partial B$, where $\Delta n\equiv n_\uparrow-n_\downarrow=(N_\uparrow - N_\downarrow)/V$. As a dimensionless quantity, we consider the normalized susceptibility
\begin{equation}\label{eq:sus_norm_FEG}
\overline{\chi}=\frac{\partial\left(\Delta n/n\right)}{\mu_B\partial\left(B/E_F[n/2]\right)},
\end{equation}
where the polarization $\Delta n /n$ is used as a more meaningful measure of magnetization. $E_F[n]=\hbar^2(6\pi^2 n)^{2/3}/{2m}$ and the total density is $n\equiv n_\uparrow+n_\downarrow=(N_\uparrow+N_\downarrow)/V$.

Here we experimetanlly simulate Pauli paramagnetism using trapped fermionic alkali atoms. The two lowest hyperfine states of $^6$Li, $\left| F=1/2, m_F=1/2 \right>\equiv\left|\uparrow\right>$ and $\left| F=1/2, m_F=-1/2 \right>\equiv\left|\downarrow\right>$, represent the two spin states of an electron. This atomic system is different from free electron systems: (1) due to the slow dipolar relaxation \cite{ketterleVarennaFermi}, the population in each spin state is conserved and  therefore, (2) an external magnetic field does not lead to any magnetization, it only shifts the energies of the two spin state (only in the case of strongly magnetic dipolar gases such as Cr the dipolar relaxation is fast, and therefore the spin populations follow an external magnetic field \cite{FattoriNatPhys2}). 

To realize the effective magnetic field we introduce density imbalance while preparing the system \cite{ShinPRL97,ShinNature451} as illustrated in Fig.~\ref{fig:paramagnetism}(b). In the grand-canonical description of a system with fixed atom number, a Lagrange multiplier (called the chemical potential) ensures the correct expectation value for the atom number.  In our case, the atom numbers in both spin states are fixed.  In a grand-canonical description, this leads to two Lagrange multipliers (the two chemical potentials for the two spin states), or to one Lagrange multiplier for the total number, and one for the population difference. It is the sum of the latter Lagrange multiplier and the external magnetic field which constitutes the effective magnetic field in our ensemble. The contribution from the external magnetic field is always canceled out by an additive component in this multiplier. For example, a balanced  gas always has zero effective magnetic field independent of the applied field. Therefore the effective magnetic field depends only on the population imbalance and not on the external magnetic field.

For a uniform Fermi gas at $T=0$, the number in each spin state determines chemical potential, $\mu_{\uparrow(\downarrow)}$ \cite{Note0} and satisfies
\begin{equation}\label{eq:cold_atom}
N_{\uparrow(\downarrow)}=\int_{0}^{\mu_{\uparrow(\downarrow)}}g(E)dE=\int_{\mp\Delta\mu/2}^{\tilde\mu}g(E\pm\Delta\mu/2)dE,
\end{equation}
where $\tilde\mu=(\mu_\uparrow+\mu_\downarrow)/2$ and $\Delta\mu=\mu_\uparrow-\mu_\downarrow$. $g(E)$ is the same as in Eq. (\ref{eq:solid_state}) if $m$ is taken to be the mass of a $^6$Li atom. Comparing with Eq.(\ref{eq:solid_state}) shows that the chemical potential difference $\Delta\mu$ is the effective magnetic field that takes the place of the magnetic field in a FEG. With this effective magnetic field, the magnetic susceptibility can be written as
\begin{equation}
\chi=\frac{\partial \Delta n}{\partial \Delta \mu}
\end{equation}

We show now how a single atomic density profile in a harmonic trap $V(\vec{r})$ can be used to determine the normalized susceptibility as a function of magnetic field. In the local density approximation, the density at each point in the trap is that of the corresponding uniform gas with modified chemical potentials $\mu_{\uparrow(\downarrow)}(\vec{r})=\mu_{\uparrow(\downarrow)}^{g}-V(\vec{r})$  \cite{GiorginiRMP80,ketterleVarennaFermi}, where  $\mu_{\uparrow(\downarrow)}^{g}$ are global chemical potentials defined for the whole trapped cloud which constrains the total number of atoms in each spin state. This concept has been used to determine the equation of state of interacting Fermi gases \cite{ShinNature451,KuScience335,HorikoshiScience327,NavonScience328}. Here we are interested in the normalized magnetic susceptibility which is the slope of the normalized density difference versus the normalized chemical potential difference \cite{Note1}. We demonstrate this procedure first using simulated density profiles (Fig.~\ref{fig:simulation}).

As $\mu_{\uparrow(\downarrow)}(\vec{r})$ varies across the cloud, the chemical potential difference $\Delta\mu(\vec{r})\equiv\mu_\uparrow(\vec{r})-\mu_\downarrow(\vec{r})=\mu_{\uparrow}^g-\mu_{\downarrow}^g$ remains constant, as shown in Fig.~\ref{fig:simulation}(b). However, the total density $n(\vec{r})=n_{\uparrow}(\vec{r})+n_{\downarrow}(\vec{r})$ changes across the cloud, and consequently the local Fermi energy $E_F[n(\vec{r})]$ as well, with $E_F[n]$ being the same as in Eq. (\ref{eq:solid_state}) if $m$ is taken to be the mass of $^6$Li. The normalized magnetic susceptibility $\overline{\chi}$ depends on the dimensionless effective magnetic field $\overline{\Delta \mu}=\Delta\mu(\vec{r})/E_F[n(\vec{r})]$, which does vary across the cloud, from small values near the center to the saturated value of one at the edges. From Fig.~\ref{fig:simulation}(c), a single set of atomic density profiles contains a range of polarizations $\overline{\Delta n}=\Delta n/n$ and normalized effective magnetic fields $\Delta\mu/E_F[n]$. The slope of the $\overline{\Delta n}$ vs. $\overline{\Delta \mu}$ yields the normalized susceptibility
\begin{equation}\label{eq:susceptibility}
\overline\chi=\partial\left(\overline{\Delta n}\right)/\partial\left(\overline{\Delta \mu}\right).
\end{equation}

The normalization of $\overline\chi$ in Eq.~(\ref{eq:susceptibility}) is chosen to ensure $\overline{\Delta n}=\overline{\Delta\mu}=1$ for a fully polarized gas. The slope near the origin, corresponding to small values of $\overline{\Delta\mu}$, gives $\overline\chi(T=0)\equiv\overline\chi_0=3/2^{4/3}\approx 1.19$ \cite{Note2}. As shown in Fig.~\ref{fig:simulation}(e), the susceptibility decreases with increasing $\overline{\Delta\mu}$ and drops to zero when the Fermi gas reaches full polarization. 

\begin{figure}[h]
  \includegraphics[scale=0.5]{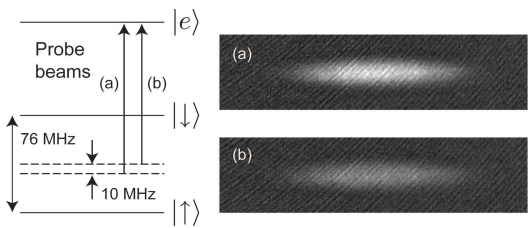}\\
  \caption{Double-shot phase contrast imaging. $\left|e\rangle\right.$ is the 2P$_{3/2}$ excited state in $^6$Li. Two successive images are taken at two different imaging frequencies, (a) and (b). Corresponding CCD images are shown on the right.}\label{fig:imaging}
\end{figure}

\section{Experiment and Results}\label{sec:experiment}

An ultracold one-component Fermi gas of $^6$Li in the state $|F=3/2,m_F=3/2\rangle$ is prepared by sympathetic cooling with bosonic $^{23}$Na atoms \cite{Jo2009,LeePRA85}.  $^6$Li atoms are then loaded into an optical dipole trap followed by radio-frequency (rf) transfer to $|F=1/2,m_F=1/2\rangle$ using an rf Landau-Zener sweep at 300 G. A superposition of the two lowest hyperfine ground states, $|F=1/2,m_F=1/2\rangle\equiv\left|\uparrow\rangle\right.$ and $|F=1/2,m_F=-1/2\rangle\equiv\left|\downarrow\rangle\right.$ is prepared by a rf sweep at 300G, and holding here for 500 ms results in an incoherent mixture. 
\begin{figure}[h]
  \includegraphics[scale=0.6]{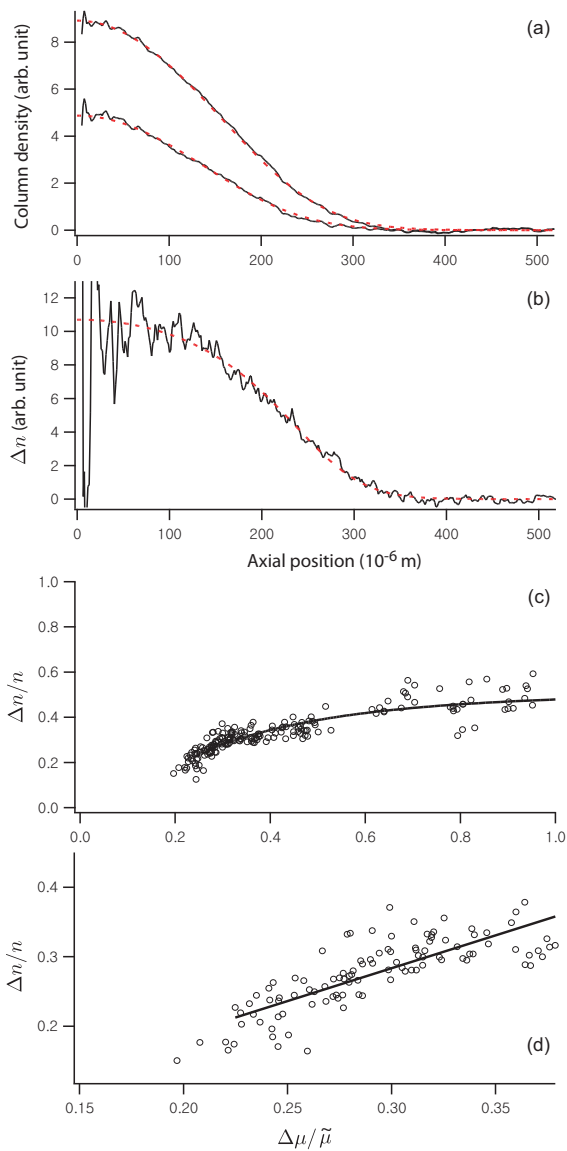}\\
  \caption{(color online) Experimental determination of $\overline\chi$. Representative profiles of (a) 2D column densities for $\left|\uparrow\rangle\right.$ (upper solid curve) and $\left|\downarrow\rangle\right.$ (lower solid curve).  Temperature and chemical potential are obtained from the fit (dotted curves) as discussed in the text. Each elliptical average is plotted versus the axial position (long axis of the ellipse). (b) 3D density difference obtained from a single double-shot image.   Dotted curve is a theoretical curve based on the chemical potentials and the temperature obtained from the 2D fit in (a). (c) The normalized 3D density difference is plotted as a function of the normalized chemical potential difference. The solid line is the theoretical prediction. (d) The linear fit at low $\overline{\Delta\mu}$ gives the low-field susceptibility $\overline\chi$=0.95(1). }\label{fig:density_profile2}
\end{figure}
Further cooling is provided by lowering the optical trapping potential which leads to radial evaporation (axial confinement is provided magnetically).  We increase the magnetic field in 200 ms to 528G, where the scattering length is zero \cite{Ohara2002}, leading to a non-interacting two-component Fermi gas. The oscillation frequencies of the final trapping potential are $\omega_y=2\pi\times35$ Hz axially and $\omega_x=\omega_z=2\pi\times 390$ Hz radially. The total population imbalance $\delta\equiv(N_\uparrow-N_\downarrow)/(N_\uparrow+N_\downarrow)$ is about 33\%. The final temperature is about 0.22$T/T_F$ where $k_B T_F=E_F[n/2]$ with $n$ being the total peak density and $k_B$ the Boltzmann constant.

To obtain density profiles of each spin state of $^6$Li, we use a double-shot phase contrast imaging technique that involves two images taken at two different imaging frequencies \cite{LeePRA85,ShinNature451} (see Fig.~\ref{fig:imaging}). One frequency is tuned between the two resonant transition frequencies for $\left|\uparrow\rangle\right.$ and $\left|\downarrow\rangle\right.$ states to the 2P$_{3/2}$ electronically excited state. The resulting phase contrast picture is a measure of the density difference $\Delta n=n_{\uparrow}-n_{\downarrow}$. To measure the density of each component, a second image is taken of the same cloud at a frequency detuned towards the $\left|\uparrow\rangle\right.$ transition by 10 MHz. The two images can be taken in quick succession because of the non-destructive nature of phase contrast imaging.

For dispersive imaging techniques such as phase contrast imaging, it is crucial to carefully focus the imaging system, since the atomic cloud refracts the probe light (in contrast to resonant absorption imaging).  One can check this by comparing in-trap profiles of $\left|\uparrow\rangle\right.$ atoms imaged at positive and negative detunings of 60 MHz and find the focal position which minimizes the difference \cite{LeePRA85}. 

Figure \ref{fig:density_profile2} shows the experimental results. The camera images provide two-dimensional (2D) column densities $n_{\rm 2D}$ which integrate the density distribution along the line of sight. To reduce the noise for the following analysis we take advantage of the symmetry of the trap and perform quadrant averaging and elliptical averaging:  First, the four quadrants around the center of the images are averaged. Second, the data is averaged along elliptical contours.  The aspect ratio of the ellipses is determined from in-trap images \cite{ShinNature451}. Temperature and global chemical potential for each component are obtained by fitting 2D densities with theoretical non-interacting density profiles. Since the temperature $T$ is small but non-negligible, data are fitted with finite-temperature Fermi gas 2D distribution in a harmonic trap given by
\begin{eqnarray}\label{eq:Finite_T_Fermi dist}
&&n_{2\rm{D}}(x,y)=\nonumber\\
&&{}-\frac{m (k_B T)^2}{2\pi\hbar^3\omega_z}{\rm Li_2}\left[-e^{\left(\mu^g-m(\omega_x^2x^2+\omega_y^2y^2)/2\right)/k_B T)}\right],
\end{eqnarray}
where ${\rm Li}_n(z)$ is the $n$th-order Polylogarithm. $z$ is the axis along the line of sight. Since the majority component has better signal-to-noise ratio, we determine the temperature and chemical potential first for this component, and keep the temperature as a fixed parameter in the fit of the minority profiles. Three-dimensional (3D) densities [Fig.~\ref{fig:density_profile2}(b)] are obtained by performing the inverse Abel transformation of the column densities \cite{ShinNature451}. The linear fit of $\overline{\Delta n}$ as a function of $\overline{\Delta\mu}$ at small $\overline{\Delta\mu}$ gives the susceptibility at finite temperature [Fig.~\ref{fig:density_profile2}(d)]. For finite but low temperatures, the susceptibility is known to vary as \cite{Pathria:StatMech}
\begin{equation}\label{eq:susceptibility_T}
\overline\chi=\overline\chi_0\left[1-\frac{\pi^2}{12}\left(\frac{T}{T_F}\right)^2\right].
\end{equation}
$T/T_F$ changes from 0.24 to 0.35 within the fitting region in  Fig.~\ref{fig:density_profile2}(d), affecting the susceptibility by about 3\%. For $T/T_F$=0.29, this theoretical calculation gives $\overline\chi=1.108$ and the experimentally obtained value is 0.95(1) in Fig.~\ref{fig:density_profile2}(d). The discrepancy between the experimental and theoretical value is most likely due to a residual dispersion effect leading to systematic uncertainty estimated to be 10-20\%  \cite{Note3,Note4}. We did not attempt to quantify this effect since this paper is mainly pedagogical and a dispersion effect depending on the imbalance as well as two detunings would be time-consuming to quantify  
  
Figure \ref{fig:density_profile2}(c) illustrates that we can easily approach the high field limit of a fully polarized gas $\overline{\Delta\mu}\approx1$.  Note that $\overline{\Delta n}$ is smaller than 1 because at finite temperature the minority component extends beyond its Thomas-Fermi radius.  In a metallic free electron gas, with a typical Fermi temperatures of 10,000 K, the high field region would require magnetic fields of 10,000 Tesla, 200 times stronger than the strongest continuous laboratory magnetic fields.

\section{Conclusions}
We have demonstrated how the \emph{ideal gas} magnetic susceptibility can be measured in ultracold Fermi gases. After considering finite-temperature effects, the experimental results are in quantitative agreement with prediction from Pauli paramagnetism. Since the local spin polarization varies across the trapped atomic cloud, a single density profile gives the susceptibility for both low and high effective magnetic fields. This method can be applied to the study of magnetic properties of  strongly interacting Fermi gases which are of current interest \cite{MeinekeNatPhys8,NascimbenePRL106,SommerNat472,SannerPRL108}.

\begin{acknowledgments}
This work was supported by the NSF and the ONR, by an AFOSR MURI on Ultracold Molecules, and by ARO grant no. W911NF-07-1-0493 with funds from the DARPA Optical Lattice Emulator program. Y.-R. Lee acknowledges support from the Samsung Scholarship. T. T. Wang acknowledges support from NSERC. We are grateful to Caleb A. Christensen for experimental assistance. 
\end{acknowledgments}

\bibliographystyle{apsrev}

\end{document}